# Crystal Growth of $Bi_2Sr_2CaCu_2O_{8+\delta}$ Whiskers from Pulverized Amorphous Precursors

*Ryo Matsumoto[a,b], Sayaka Yamamoto[a,b], Yoshihiko Takano[a,b], Hiromi Tanaka[c],*

*[a]International Center for Materials Nanoarchitectonics (MANA),
National Institute for Materials Science, 1-2-1 Sengen, Tsukuba, Ibaraki 305-0047, Japan*
*[b]University of Tsukuba, 1-1-1 Tennodai, Tsukuba, Ibaraki 305-8577, Japan*
*[c]National Institute of Technology, Yonago College, 4448 Hikona, Yonago, Tottori 683-8502, Japan*

**Abstract**

High-transition temperature superconducting whiskers of $Bi_2Sr_2CaCu_2O_{8+\delta}$ were successfully grown from pulverized amorphous precursors. The obtained whiskers revealed a typical composition and diffraction patterns of a superconducting Bi-2212 phase. The whiskers from tiny precursors exhibit a spiral feature, suggesting a contribution from the growth mechanism of both vapor-liquid-solid and compressive stress models. The proposed method provides longer whiskers against the growth period, compared with that from conventional root-growth method.

# 1. Introduction

Since a discovery of superconducting $Bi_2Sr_2Ca_{n-1}Cu_nO_{2n+4+\delta}$ with transition temperature ($T_c$) of 40 K ($n$=1), 80 K ($n$=2), and 110 K ($n$=3) [1-2], developments for wire applications have been continued to transport electricity without any energy loss [3-5]. A critical current density ($J_c$), which is limited value of current in the superconductors, is bottleneck for a practical use of such superconducting materials for the wire application [6]. A growing of strong pinning centers is most important strategy to enhance the $J_c$ in the superconducting wires [7-8]. Especially, an investigation for the pinning effects for an intragrain $J_c$ is required to understand an intrinsic property of the superconducting materials.

A superconducting whisker crystal is one of the promising materials to investigate the intragrain $J_c$ due to their superior crystalline nature excluding a pinning effect from defects in the crystal. Recent studies on $Bi_2Sr_2CaCu_2O_{8+\delta}$ (Bi-2212) whiskers revealed a drastic enhance of intragrain $J_c$ up to $2\times10^5$ A/cm$^2$ beyond a practical criterion for the wire application by introduced pillar-shaped nanocrystalline domains [9]. A partial substitution for $Mg^{2+}$ ion into $Ca^{2+}$ site improved a $J_c$ anisotropy against an applied magnetic field in Bi-2212 whiskers [10]. These knowledges are quite important how to enhance the electrical transport properties of not only Bi-2212 cuprate but also the wide superconducting materials.

One of the most conventional procedures to obtain the Bi-2212 whiskers are $Al_2O_3$-seeded glassy quenched platelet (ASGQP) method [11]. Although the ASGQP method maximumly provides ~10 mm length of the Bi-2212 whiskers, the dimension of the precursors has not been optimized yet. In this study, we proposed a use of pulverized precursors in the whisker growth. The maximum crystal-size and growth speed of the Bi-2212 whiskers were improved by using the size-controlled precursors.

# 2. Experimental procedures

The precursors of the Bi-2212 whiskers were prepared by using conventional ASGQP method, as described the literatures [11]. The obtained precursors were cut into a dimension of approximately $4 \times 6 \times 0.7$ mm and then placed onto an $Al_2O_3$ boat as shown in Fig. 1(a). On the other hand, we also prepared roughly ground precursors to confirm the size-effect for the whisker growth. The obtained precursors vie conventional ASGQP method were ground into rough tips with dimension of approximately $1 \times 1 \times 0.7$ mm and then spread onto the $Al_2O_3$ boat as shown in Fig. 1(b). The $Al_2O_3$ boats with two kinds of precursors were annealed in a tube furnace with same conditions of the temperature: 850-900°C, oxygen flow amount: 120 ml/min, and period: 24-168 hours. The dimension and compositional ratio of the obtained whisker were evaluated by a scanning electron microscopy (SEM) equipped with an energy dispersive spectrometry (EDX) using a TM3000 (Hitachi High-Technologies). The crystal structure was investigated by an X-ray diffraction (XRD) measurement using Ultima IV (Rigaku) with Cu K$\alpha$ radiation ($\lambda$ = 1.5418 Å).

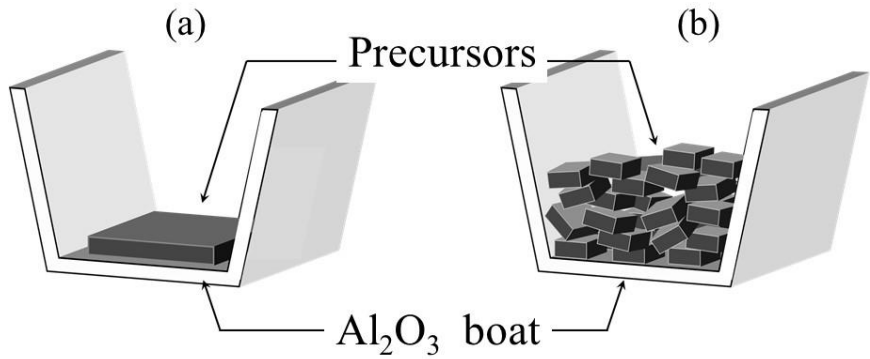

**Figure 1. Schematic images of the precursor preparation for (a) conventional ASGQP method and (b) pulverized ASGQP method.**

**3. Results and discussion**

Figure 2 (a) shows an optical microscope image of the obtained whiskers from the pulverized precursors. A lot of whiskers grown from the precursors with enough length. The inside of the $Al_2O_3$ boat became partially yellow, indicating a vaporization of the starting materials of the whisker during the crystal growth. The SEM observation, as shown in Fig. 2 (b), revealed that some of the obtained whiskers exhibit spiral shape despite the fact that the whiskers from the conventional method are straight needle-like crystal. Generally speaking, the Bi-2212 whisker grows from their root [12]. However, the growth mechanism of coiled whiskers in some materials, for example, $Si_3N_4$ [13], is considered to be a vapor-liquid-solid (VLS) process. The spiral feature of our obtained Bi-2212 whiskers from the pulverized precursors seems to be the coiled whiskers from VLS process.

Figure 2 (c) shows a typical XRD pattern in the obtained whiskers from the pulverized precursors. The (00$l$) XRD peaks from a typical Bi-2212 superconducting whisker were observed in the obtained whiskers [10]. The EDX analysis yields that the chemical composition of the obtained whiskers was Bi : Sr : Ca : Cu = 2 : 1.5 : 1.5 : 2 which exhibits a Sr poor and Ca rich feature reflecting the nominal component of the precursor. According to the XRD and EDX measurements, the obtained whiskers from the pulverized precursors could be considered as typical Bi-2212 whiskers.

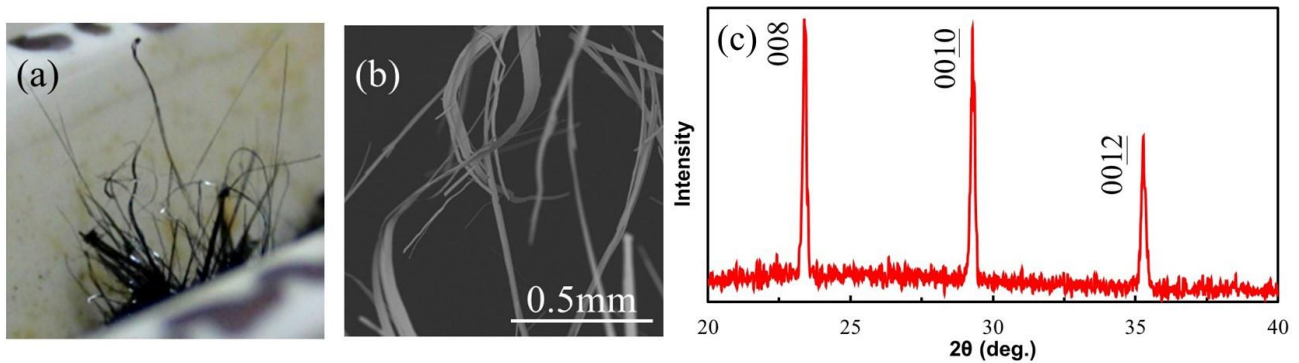

**Figure 2. (a) Optical microscope image and (b) SEM image of the obtained whiskers from the pulverized precursors. (c) Typical XRD pattern in the obtained whiskers from the pulverized precursors.**

The Bi-2212 whiskers were grown from conventional ASGQP precursors and the proposed method using the pulverized precursors under various growth period up to 170 hours, and then the maximum length among the grown whiskers were evaluated by using SEM observation. Figure 3 shows a comparison for a maximum length of the obtained Bi-2212 whiskers at each growth period from conventional ASGQP precursors and the proposed method using the pulverized precursors. The growing curve of the conventional ASGQP method was tended to saturate from 50 hours and provided ~5 mm whiskers maximumly. On the other hand, our proposed method using the pulverized precursors provided ~11 mm length of the Bi-2212 whiskers maximumly without the saturation of the growth rate. Most of the whiskers exhibited the spiral feature, indicating a possibility for a contribution of VLS process to the origin of the increase of the crystal-size. Here it should be noted that some possibilities for a growth mechanism of the Bi-2212 whiskers in our proposed method. According to the literatures [14-16], the driving force on most of metal and oxide whiskers is considered relating an internal stress in their precursors. This growth model suggest that the whiskers grow during a reduction of the internal stress in the precursors under annealing condition. The relative internal stress against the size of precursors became large in the case of our proposed method using pulverized precursors, compared to that of the conventional ASGQP method. These facts suggest that the VLS-like growth and the internal stress-driven growth were accelerated by the increase of the vaporized gas and decrease of the precursor size, respectively, in the proposed method. As a future work, more positive use of these components for the whisker growth is expected to further understand the growth mechanism and how to obtain more longer crystal of the Bi-2212 whiskers.

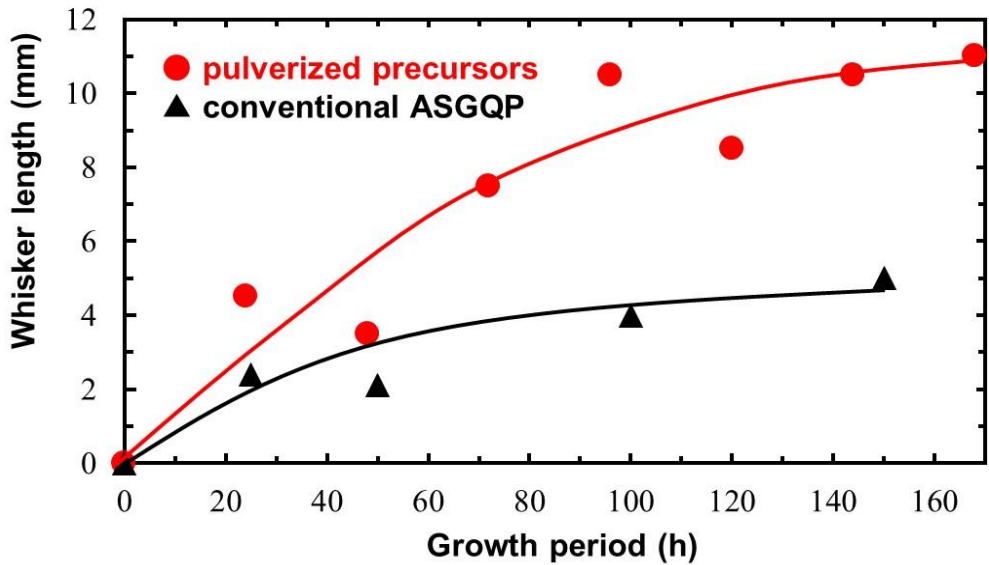

**Figure 3. Comparison for a maximum length of the obtained Bi-2212 whiskers at each growth period from conventional ASGQP precursors and the proposed method using the pulverized precursors.**

## 4. Conclusion

We focused on the size of the precursors to obtain the longer Bi-2212 superconducting whiskers. As the expectation, the growth rate of the whiskers was drastically improved in our proposed method using the pulverized precursors. The obtained whiskers exhibited the spiral feature, known as the contribution from the VLS growth. On the other hand, the compressive stress, which is considered as the driving force of the most of metal whiskers, is maybe emphasized in this method due to the small size of the precursors. These components possibly provide the longer whisker in our method. The further investigations are expected for the evidence of the improvement of the growth rate, for example, quantitative evaluation of the compressive stress.

**Acknowledgment**

The authors thank Mr. K. Kamimoto and Ms. K. Maeda for strong supports regarding to the sample preparations and measurements. This work was partly supported by JSPS KAKENHI Grant Number JP17J05926.